\documentstyle[12pt,amssymb]{article}
\begin{document}
\begin{titlepage}
\begin{center}
{\Large\bf Fast Quantum Maps}

\vskip1.5truecm
G. G. Athanasiu$^{*}$\footnote{e-mail: athanasi@physics.uch.gr}, 
E. G. Floratos$^{**}$\footnote{e-mail: manolis@timaios.nrcps.ariadne-t.gr.
On leave of absence from Physics Department, University of Crete.}
 and S. Nicolis$^{***}$\footnote{e-mail: nicolis@celfi.phys.univ-tours.fr}

\vskip1truecm 

$^{*}${\sl Physics Department, University of Crete\\
{\rm and}\\
F.O.R.T.H., Heraklion, Crete, Greece}

\vskip1truecm 

$^{**}${\sl I. N. P. , NRCPS ``Demokritos''\\
15310 Aghia Paraskevi, Athens, Greece} 

\vskip1truecm

$^{***}${\sl CNRS--Laboratoire de Math\'ematiques et Physique Th\'eorique (UPRES A 6083)\\
 Universit\'e de Tours, Parc Grandmont, 37200 Tours, France}

\end{center}

\vskip1truecm

\begin{abstract}
{\small 
We develop number theoretic tools that allow to perform computations
relevant for the quantum mechanics over finite fields of arbitrary, odd size, 
with the same speedup that is enjoyed by the Fast Fourier Transform. }
\end{abstract}

\end{titlepage}

Recent developments in non-perturbative string theory,the discovery of 
D-branes~\cite{polchinski} and their dynamics,revealed a new sector of
heavy solitonic objects through which distances below string scales
can be probed in the weak coupling regime(non-relativistic limit)\cite{shenker}.
It seems that the most fundamental of these solitonic objects, the D0 branes,
have low energy effective lagrangian the 1-dimensional reduction of
a 10d supersymmetric YM system,the so called SUSY YM quantum mechanics.
Indeed a stack of N  D0 branes has $SU(N)$ SUSY YM quantum mechanics as its 
low energy effective lagrangian~\cite{witten,shortdistance} and the target space collective 
coordinates of the D0 branes become $N$ by $N$ hermitian matrices functions of
time[YM gauge potentials] and their SUSY partners.This implies that a non-
commutative geometry setting is emerging for the description of the
dynamics of D0 branes\cite{douglas}.

All the above has been lifted to the level of a candidate for the M-theory,
(the theory which presumably unifies all known string theories),   
the famous by now M(atrix) Theory~\cite{bfss}
Curiously enough,the above picture resembles the (many years old)
$SU(N)$ truncation\cite{nicolai} of the excitations of the
supermembrane theory in 11 dimensions\cite{townsend} in analogy with the
bosonic membrane $SU(N)$ truncation\cite{hoppe,floratos}.

More close to the D0 picture comes the work of \cite{floratos},where
the discretized membrane and its noncommutative geometry,Finite Quantum 
Mechanics,was introduced, as a consistent truncation of the bosonic membrane
and its dynamical symmetry that of the area preserving diffeomorphism group.
In this discretized version of the membrane the elementary excitations-
degrees of freedom were assumed to be one particle states living on the
membrane, like particles in discrete phase space 
${\Bbb Z}_N\times {\Bbb Z}_N$.
A physical
analog system of these elementary excitations was proposed to be the
Quantum Hall effect of one electron on a magnetic lattice of rational
magnetic flux per plaquette.The Hilbert space of these elementary excitations
is finite dimensional and the quantum mechanics of linear quantum maps
was further developed in refs\cite{bi,afetal,afn}.
The $SU(N)$ matrices of the YM quantum mechanics can be thought as coherent
states of such elementary excitations.The difference with the above mentioned
model of elementary excitations of the membrane is that time is also discrete
and the motion of these excitations is typically random and chaotic--
a fact which
 at the quantum level is translated into extended, random wave functions for
typical eigenstates.

Although we are far from a realistic scenario for the
role of these elementary excitations for the quantum dynamics of the 
SUSY $SU(N)$ Quantum Mechanics we believe that further technical developments
are necessary in order to acquire better understanding of the situation.

On a more mathematical side Finite Quantum Mechanics [FQM] has been developed
so far using representation theory of the modular group $SL(2,{\Bbb Z}_N)$,the
linear canonical transormation group of the elementary excitations,for values
of $N$,prime or powers of primes. In this note we treat the case
of general odd integers $N$,using prime decomposition and the Chinese remainder
theorem for the modular group and its representations.The case of integers
$N=2^n$ and general integers will be dealt with elsewhere.

An immediate, practical, consequence of our work is the 
possibility to extend the Fast Fourier transform for any odd $N$ to the 
metaplectic representation of the modular group $SL(2,{\Bbb Z}_N)$  

We now recall the basic features of  FQM. 

The torus phase space has been the simplest prototype for studying classical
and quantum chaos~\cite{qc2,qc3,qc4,qc5}. Discrete 
elements of $SL(2,{\Bbb R})$, i.e. elements of the modular group 
$SL(2,{\Bbb Z})$, are studied on discretizations of the torus with rational 
coordinates of the same denominator $l$, $(q,p)=(n_1/l,n_2/l)\in\Gamma,\,
n_1,n_2,l\in{\Bbb Z}$ and their periodic trajectories mod 1 are examined studying 
the periods of elements ${\cal A}\in SL(2,{\Bbb Z})$ mod $l$. The action mod 1 
becomes mod $l$ on an equivalent torus, $(n_1,n_2)\in l\Gamma$. The classical motion of such discrete dynamical systems is usually ``maximally''
disconnected and chaotic~\cite{qc3,qc5}.

FQM is the  quantization of these discrete linear maps and the corresponding 
one-time-step
evolution operators $U({\cal A})$ are $l\times l$ unitary matrices called 
{\em quantum maps}. In the literature~\cite{qc4,qc5}
these maps are determined semi-classically. In ref.~\cite{bi,afetal} 
the exact quantization of $SL(2,{\Bbb F}_p)$, 
where ${\Bbb F}_{p}$ is the simplest finite field of $p$ elements with 
$p$ a prime number was studied in detail. In ref.~\cite{afn} these results 
were extended to powers of primes, $p^n$--the group is then $SL(2,{\Bbb Z}_{p^n})$.

The Hilbert space ${\Bbb H}_{\Gamma}$ of the wave functions on the torus 
$\Gamma={\Bbb C}/{\Bbb L}$ of complex modulus $\tau=\tau_1+{\mathrm i}\tau_2$,
where ${\Bbb L}$ is the integer lattice, ${\Bbb L}=\{m_1+\tau m_2|(m_1,m_2)\in 
{\Bbb Z}\times{\Bbb Z}\}$ is defined as the space of functions of complex
argument $z=x+{\mathrm i}y$ 
\begin{equation}
f(z)=\sum_{n\in{\Bbb Z}}c_n{\mathrm e}^{{\mathrm i}\pi n^2\tau+2\pi{\mathrm i}nz}
\end{equation}
with norm~\cite{bi}
\begin{equation}
||f||^2=\int {\mathrm e}^{-2\pi y^2/\tau_2}|f(z)|^2dxdy,\,\,\,\tau_2>0 
\end{equation}
Consider the  subspace ${\Bbb H}_{l}(\Gamma)$ of ${\Bbb H}_{\Gamma}$ 
with periodic Fourier coefficients $\{c_n\}_{n\in{\Bbb Z}}$ of period $l$
\begin{equation}
\label{fourier}
c_n=c_{n+l}\,\,\,n\in {\Bbb Z},\,\,l\in{\Bbb N}
\end{equation}
The space   ${\Bbb H}_{l}(\Gamma)$ is $l$-dimensional and there is a discrete
Heisenberg group~\cite{Weyl}, 
with generators ${\cal S}_{1/l}$ and ${\cal T}_{1}$ 
acting as~\cite{cartier,mumford}
\begin{eqnarray}
\label{STl}
({\cal S}_{1/l}f)(z) &=& \sum_{n\in{\Bbb Z}}c_n
{\rm e}^{2\pi{\rm i}n/l}{\rm e}^{2\pi{\rm i}nz+\pi{\rm i}n^2\tau}\nonumber\\
 & & \nonumber\\
({\cal T}_{1}f)(z) &=& \sum_{n\in{\Bbb Z}}c_{n-1}
{\rm e}^{2\pi{\rm i}nz+\pi{\rm i}n^2\tau},\,\,\,c_n\in{\Bbb C}
\end{eqnarray}
On the $l$-dimensional subspace of vectors $(c_1,\ldots,c_l)$ the two 
generators are represented by
\begin{eqnarray}
\label{gener}
({\cal S}_{1/l})_{n_1,n_2}=Q_{n_1,n_2}=\omega^{(n_1-1)}\delta_{n_1,n_2}
\nonumber\\
\nonumber   \\
({\cal T}_{1})_{n_1,n_2}=P_{n_1,n_2}=\delta_{n_1-1,n_2}
\end{eqnarray}
with $\omega=\exp(2\pi{\rm i}/l)$. The Weyl relation becomes 
\begin{equation}
\label{Weyl}
QP=\omega PQ
\end{equation}
and the Heisenberg group elements are
\begin{equation}
\label{heisel}
{\cal J}_{r,s}=\omega^{r\cdot s/2}P^rQ^s
\end{equation}

The generators ${\cal J}_{r,s}$ satisfy the following composition law
\begin{equation}
{\cal J}_{ r,s}{\cal J}_{r',s'}=\omega^{\displaystyle (r's-s'r)/2}{\cal J}_{\displaystyle r+r',s+s'}
\end{equation}
and the ``commutation'' relations
\begin{equation}
{\cal J}_{r,s}{\cal J}_{r',s'}=\omega^{\displaystyle r's-s'r}
{\cal J}_{r',s'}{\cal J}_{r,s}
\end{equation}
The metaplectic representation of $SL(2,{\Bbb Z}_l)$ is defined by the relation
\begin{equation}
U^{-1}({\cal A}){\cal J}_{r,s}U({\cal A})={\cal J}_{(r,s){\cal A}}
\end{equation}
where ${\cal A}$ is an element of $SL(2,{\Bbb Z}_l)$.
In the literature 
the metaplectic representation of $SL(2,{\Bbb Z}_l)$, (the group of $2\times 2$,
integer valued 
matrices mod $l$), is known for 
$l=p^n$~\cite{tanaka}\footnote{The representation theory of the symplectic 
group $SL(2,{\Bbb F}_{p^n})$, over the finite field ${\Bbb F}_{p^n}$.
 may be found in ref.~\cite{tanaka2}.}

The Weyl-Fourier form of $U({\cal A})$ is~\cite{afn}
\begin{equation}
\label{weylfour}
U({\cal A})=\frac{\sigma(1)\sigma(\delta)}{p^n}
\sum_{r,s=0}^{p^n-1} {\rm e}^{\frac{2\pi{\rm i}}{p^n}
[br^2+(d-a)rs-cs^2]/2\delta}
{\cal J}_{r,s}
\end{equation}
where 
\begin{eqnarray}
\label{sl2defs}
{\cal A} &=&\left(\begin{array}{cc} a & b\\c &d\end{array}\right)\in SL(2,{\Bbb Z}_{p^n}),\,\,\,\delta=2-a-d\nonumber\\
 & & \nonumber\\
\sigma(x)&=&\frac{1}{\sqrt{p^n}}\sum_{r=0}^{p^n-1}\omega^{xr^2}
\end{eqnarray}

All the operations in the exponent are carried out mod $p^n$.
If $\delta\equiv 0$ mod $p^n$ we use the trick
\begin{equation}
\label{trick}
\left(\begin{array}{cc}a & b\\c & d\\ \end{array}\right) = 
\left(\begin{array}{cc} 0 & 1 \\ -1 & 0\\ \end{array}\right)
\left(\begin{array}{cc} -c & -d \\ a & b\\ \end{array}\right)
\end{equation}
and the fact that $U({\cal A})$ is a representation (cf. ref.~\cite{afn} and below).

We shall now work out some technical details 
of the representation theory of $SL(2,{\Bbb Z}_{N})$, that we need for the 
factorization of the Heisenberg group and of the metaplectic representation. 

We start with the Chinese remainder theorem for numbers\cite{sigproc1}.
 Let $N\in{\Bbb Z}$ be a non-prime, that may be written as a product of two co-prime factors, $N_1$ and $N_2$, {\em viz.} $N=N_1N_2$. We denote by $N{\Bbb Z}$ the set of all multiples of $N$. Then any $r\in {\Bbb Z}/N{\Bbb Z}$ may be written uniquely as
$$
r=r_1m_1n_1+r_2m_2n_2
$$
where
$r_1\equiv r{\mathrm mod}N_1$, $r_2\equiv r{\mathrm mod}N_2$, $m_1=N/N_1$,$m_2=N/N_2$, $n_1=m_1^{-1}{\mathrm mod}N_1$,$n_2=m_2^{-1}{\mathrm mod}N_2$.

In other words, we may establish a one-to-one correspondance between the number $r$ and the 2-tuple $(r_1,r_2)$. This last defines the {\em Sino representation} of $r$. 
This 2-tuple may be promoted to a bona fide element of a set ${\cal V}_N^{(2)}$, whose elements have the following properties,
for any two numbers $r,r'\in {\Bbb Z}/N{\Bbb Z}$:
\begin{itemize}
\item  
$$r\times r'\leftrightarrow 
(r_1r_1^{'}{\mathrm mod}N_1,r_2r_2^{'}{\mathrm mod} N_2)
$$
and
\item 
$$r+r'\leftrightarrow (r_1+r_1^{'}{\mathrm mod} N_1,r_2+r_2{\mathrm mod}N_2)$$
\end{itemize}
It is immediate to generalize this result to the case where $N=N_1\times N_2
\times\cdots N_k$ where all pairs of factors are co-prime. The decomposition
reads
$$
r=r_1m_1n_1+r_2m_2n_2+\cdots + r_km_kn_k
$$
where $m_i=N/N_i$,$n_i\equiv m_i^{-1}{\mathrm mod}N_i$ and one may similarly 
establish a one-to-one correspondance between $r$ and the $k-$tuple 
$(r_1,r_2,\ldots,r_k)$, element of the set ${\cal V}_{N}^{(k)}$.
Furthermore note that 
 ${\cal V}_N$ has the property
$$
{\cal V}_N\leftrightarrow {\cal V}_{N_1}\otimes{\cal V}_{N_2}\otimes\cdots\otimes {\cal V}_{N_k}
$$
Using these relations it is now possible to establish that
\begin{equation}
SL(2,{\Bbb Z}_N)=SL(2,{\Bbb Z}_{N_1})\times SL(2,{\Bbb Z}_{N_2})\times\cdots\times SL(2,{\Bbb Z}_{N_k})
\end{equation}
Indeed, consider the case $k=2$ and an element of $SL(2,{\Bbb Z}_N)$ of the form
\begin{equation}
{\cal A}=\left(\begin{array}{cc}
a & b\\ c & d\\ \end{array}
\right)\leftrightarrow
\left(\begin{array}{cc}
(a_1,a_2) & (b_1,b_2) \\ (c_1,c_2) & (d_1,d_2) \\ \end{array}\right)
\end{equation}
It will now be shown that this element is an element of the set 
$SL(2,{\Bbb Z}_{N_1})\times SL(2,{\Bbb Z}_{N_2})$
Consider a generic element of $SL(2,{\Bbb Z}_{N_1})$. It may be written as
\begin{equation}
{\cal A}_1=\left(\begin{array}{cc}
(a_1,1) & (b_1,0) \\ (c_1,0) & (d_1,1) \\ 
\end{array}\right)
\end{equation}
Take now a generic element of $SL(2,{\Bbb Z}_{N_2})$, that may be written
\begin{equation}
{\cal A}_2=\left(\begin{array}{cc}
(1,a_2) & (0,b_2) \\ (0,c_2) & (1,d_2) \\ \end{array}\right)
\end{equation}
It is straightforward to check that ${\cal A}_1\cdot {\cal A}_2={\cal A}$. Using the Chinese 
remainder theorem for numbers we know that the decomposition is unique.

Let us close with the remark that the matrices of the form
\begin{equation}
\left(\begin{array}{cc} a & b \\ -b & a \\ \end{array}\right)
\end{equation}
with $a^2+b^2\equiv 1{\mathrm mod} N$ generate the group $O_2(N)\lhd SL(2,{\Bbb Z}_{N})$. Once more it is possible to show that 
\begin{equation}
O_2(N)=O_2(N_1)\times O_2(N_2)
\end{equation}
{\em viz.}
\begin{equation}
\left(\begin{array}{cc} a & b \\ -b & a \\ \end{array}\right)=
\left(\begin{array}{cc} (a_1,1) & (b_1,0) \\ (-b_1,0) & (a_1,1) \\ 
\end{array}\right)
\left(\begin{array}{cc} (1,a_2) & (0, b_2) \\ (0,-b_2) & (1,a_2) \\ 
\end{array}\right)
\end{equation}
In the following we discuss the implications of the factorization of 
$SL(2,{\Bbb Z}_N)$ (cf. previous discussion) for the 
metaplectic representation. We begin with the factorization of the Heisenberg
group ${\frak h}(N=N_1N_2)$ (cf. the work of J. Schwinger in~\cite{Weyl}).

Indeed, using the Chinese 
remainder theorem and the commutation relations of the ${\cal J}_{r,s}$, 
we find
$$
\begin{array}{c}
{\cal J}_{r,s}={\cal J}_{\displaystyle r_1m_1n_1+r_2m_2n_2,s_1m_1n_1+s_2m_2n_2}=\\
\\
{\cal J}_{\displaystyle r_1m_1n_1,s_1m_1n_1}
{\cal J}_{\displaystyle r_2m_2n_2,s_2m_2n_2}=\\
\\
{\cal J}_{\displaystyle r_2m_2n_2,s_2m_2n_2}
{\cal J}_{\displaystyle r_1m_1n_1,s_1m_1n_1}\\
\end{array}
$$
since the extra phase factor equals unity. 

We shall now use the factorization properties of the 
Heisenberg group generators, ${\cal J}_{r,s}$ 
to obtain the decomposition of the unitary operator $U({\cal A})$,
${\cal A}\in\,SL(2,{\Bbb Z}_N)$, that governs the time evolution of the quantum system in the case in hand. To do this we recall that the evolution of the generators ${\cal J}_{r,s}$ is given by
\begin{equation}
U^{-1}({\cal A}){\cal J}_{r,s}U({\cal A})={\cal J}_{(r,s){\cal A}}
\end{equation}
If $N=N_1N_2$, then ${\cal A}={\cal A}_1\cdot {\cal A}_2$, with
${\cal A}_1~\in~SL(2,{\Bbb Z}_{N_1})$ and ${\cal A}_2~\in~SL(2,{\Bbb Z}_{N_2})$. We shall show that $U({\cal A})=U({\cal A}_1)\cdot U({\cal A}_2)$.

{\em Proof:} $U({\cal A})$ may be written as a linear combination of the generators ${\cal J}_{r,s}$ as
\begin{equation}
U({\cal A})=\frac{\sigma(1)\sigma(\delta)}{N}\sum_{\displaystyle r,s=0}^{N-1}
\omega^{\displaystyle (br^2+(d-a)rs-cs^2)/(2\delta)}{\cal J}_{r,s}
\end{equation} 
where 
$$
{\cal A}=\left(\begin{array}{cc}
a & b\\ c & d\\ \end{array}
\right)
$$
Using the Sino representation, $xr^2=(x_1,x_2)(r_1^2,r_2^2)=(x_1r_1^2,x_2r_2^2)=x_1r_1^2n_1m_1+x_2r_2^2n_2m_2$ and the double sum is seen to split into the 
product of two sums
\begin{equation}
\begin{array}{c}
\displaystyle\frac{1}{N}\sum_{r=0}^{N-1}\omega^{xr^2}=\frac{1}{N_1N_2}\sum_{r_1,r_2=0}^{N_1-1,N_2-1}e^{\displaystyle 2\pi{\mathrm i}(x_1r_1^2n_1m_1+x_2r_2^2n_2m_2)/(N_1N_2)}=\\ \displaystyle
\left(\frac{1}{N_1}\sum_{r_1=0}^{N_1-1}e^{\displaystyle 2\pi {\mathrm i} m_1x_1r_1^2
/N_1}\right)\times
\left(\frac{1}{N_2}\sum_{r_1=0}^{N_2-1}e^{\displaystyle 2\pi {\mathrm i} m_2x_2r_2^2
/N_2}\right)\\
\end{array}
\end{equation}
which leads to the relation
\begin{equation}
\sigma(x)=\sigma(m_1x_1)\sigma(m_2x_2)
\end{equation}
This takes care of the prefactor. The phase is re-arranged as follows: 
$$
\begin{array}{c}
\phi\equiv (br^2+(d-a)rs-cs^2)/(2\delta)=\\
\left( (b_1r_1^2+(d_1-a_1)r_1s_1-c_1s_1)/(2\delta_1) \right)m_1n_1 + \\
\left( (b_2r_2^2+(d_2-a_2)r_2s_2-c_2s_2)/(2\delta_2) \right)m_2n_2 = \\
\phi_1N_2m_1+\phi_2N_1m_2\\
\end{array}
$$
The upshot of this is that $U(g)$ may be rewritten as
$$
U({\cal A})=\frac{\sigma(m_1\delta_1)\sigma(m_2\delta_2)}{N_1N_2}
\sum_{r_1,r_2=0}^{N_1-1,N_2-1}\omega_{N_1}^{m_1\phi_1}\omega_{N_2}^{m_2\phi_2}
{\cal J}_{r_1m_1n_1,s_1,m_1,n_1}{\cal J}_{r_2m_2n_2,s_2m_2n_2}=
$$
$$
\frac{\sigma(m_1\delta_1)}{N_1}\sum_{r_1=0}^{N_1-1}\omega_{N_1}^{m_1\phi_1}
{\cal J}_{r_1m_1n_1,s_1m_1n_1}
\frac{\sigma(m_2\delta_2)}{N_2}\sum_{r_2=0}^{N_2-1}\omega_{N_2}^{m_2\phi_2}
{\cal J}_{r_2m_2n_2,s_2m_2n_2}=U({\cal A}_1)\cdot U({\cal A}_2) 
$$
As a consequence $U({\cal A}_1)$ and $U({\cal A}_2)$ commute. Now we establish an 
isomorphism between $U({\cal A}_1)$ and $U({\cal A}_2)$ with $U_1({\cal A}_1)$ and $U_2({\cal A}_2)$,
where $U_1$ and $U_2$ are the metaplectic representations of dimension $N_1$ 
and $N_2$ respectively. Indeed we shall exhibit a permutation matrix $R$
with the properties
\begin{itemize}
\item $$ RPR^{\mathrm T}=P_1\otimes P_2 $$
\item $$ RQR^{\mathrm T}=Q_1\otimes Q_2 $$
\item $$ R{\cal J}_{r,s}R^{\mathrm T}={\cal J}_{r_1,s_1}\otimes {\cal J}_{r_2,s_2} $$
\item $$ RU(g)R^{\mathrm T}=U_1(g_1)\otimes U_2(g_2) $$
\end{itemize} 
where $P$ and $Q$ ($P_i$, $Q_i$,$i=1,2$) are the generators of the Heisenberg
group ${\frak h}(N)$ (resp. ${\frak h}(N_i)$ ).

It is enough to prove that the matrix $R$ has the property
\begin{equation}
\label{rmatrixprop}
R{\bf e}_k={\bf e}_{k_1}\otimes{\bf e}_{k_2}
\end{equation}
where ${\bf e}_k$ (${\bf e}_{k_i}$,$i=1,2$) are the eigenvectors of $P$ (resp.
$P_i$,$i=1,2$) and the indices run as $k=1,\ldots,N$,$k_i=1,\ldots,N_i$,$i=1,2$.
The other properties indeed are consequences of this. 

We start with an explicit form for the eigenvectors of $P$ 
\begin{equation}
{\bf e}_{k,l}=\frac{\omega^{k(l-1)}}{\sqrt{N}},\,\,\,k,l=1,\ldots,N
\end{equation}
Using the Sino representation we find
\begin{equation}
\label{sinorep}
\frac{\omega_N^{k_1(j_1-1)m_1n_1+k_2(j_2-1)m_2n_2}}{\sqrt{N_1N_2}}=
\frac{\omega_{N_1}^{k_1(j_1-1)m_1}}{\sqrt{N_1}}
\frac{\omega_{N_2}^{k_2(j_2-1)m_2}}{\sqrt{N_2}}
\end{equation}
In order to construct the matrix $R$, we compare the rhs of eqs.(\ref{rmatrixprop},\ref{sinorep}). The indices $j_1$ and $j_2$, in eq.(\ref{sinorep}),
from the Sino decomposition of $j$, run from 1 to $N_1$ (resp. $N_2$) and the
corresponding values of $j$ fill up a $N_1\times N_2$ array. On the other hand,
the rhs of eq.(\ref{rmatrixprop}) defines, through the tensor product, a 
decomposition of $j$ into indices $j_1$ and $j_2$ and defines another $N_1\times N_2$ array, that  is related to the previous one by a permutation matrix,
namely $R$. Specifically, if we denote the Sino decomposition array by
\begin{equation}
\{j_1,j_2\}
\end{equation}
the matrix $R$, with rows indexed by $i$ and columns indexed by $j$, has 
elements equal to 1 when $\{j_1,j_2\}=i$ and zero otherwise. This construction
is straightforwardly generalized to more than two co-prime factors of $N$
(cf. the book by Schroeder in \cite{sigproc1}). The stationary eigenvalue problem for the unitary evolution operator $U$ is reduced to that corresponding 
to each individual factor of the tensor decomposition of the matrix $U$ (e.g.
in the case where $N_1$ and $N_2$ are powers of primes of the type $4k+1$ 
explicit expressions are known for the eigenvectors and eigenvalues\cite{afetal,afn}).

The matrix $R$ is used in 
(classical) Fast Fourier algorithms to reduce the number of operations from
$O(N^2)$ to $O(N\log N)$\cite{sigproc1}).

By construction, therefore, the property in eq.(\ref{rmatrixprop}) holds
and this implies the decomposition of the $P$ operator. The decomposition
of the $Q$ operator follows immediately, since it is diagonal in this basis 
and the operations may be carried out element by element. From these both the 
decomposition of the ${\cal J}_{r,s}$ and $U(g)$ follow since (as may be checked) $RR^{\mathrm T}=I$. 

We close with the following remarks. The case $N=2^n$ cannot be studied by the 
methods developed here and new ideas are required. 
This case is, of 
course, particularly interesting for computational reasons. Indeed, all existing Fast Fourier algorithms are given for $N=2^n$. For powers of primes a similar speedup of operations may also be obtained (cf. Schroeder in \cite{sigproc1}). On the other hand, what we have achieved here is the construction of fast
algorithms for {\em any odd} $N$ and for {\em any quantum map} that is a 
metaplectic representation of $SL(2,{\Bbb Z}_N)$.

It would be interesting to implement such maps by quantum gates, as already 
proposed for the quantum Fourier transform~\cite{shor}.

\end{document}